\documentclass[3p,twocolumn]{elsarticle}

\usepackage{lineno,hyperref,amsmath,amssymb}
\modulolinenumbers[10]

\journal{New Astronomy Reviews}

\biboptions{authoryear}

\newcommand{\kepler}{\textit{Kepler}}
\newcommand{\ikt}{{\it Kepler}}
\newcommand{\ik}{{\it Kepler~}}
\newcommand{\aj}{AJ}
\newcommand{\apj}{ApJ}
\newcommand{\apjl}{ApJL}
\newcommand{\apjs}{ApJS}
\newcommand{\mnras}{MNRAS}
\newcommand{\aap}{A\&A}
\newcommand{\nat}{Nature}
\newcommand{\araa}{ARA\&A}

\begin{document}

\begin{frontmatter}

\title{The discovery and legacy of \kepler 's multi-transiting planetary systems}

\author{Jason H. Steffen}
\address{University of Nevada, 4505 S. Maryland Pkwy, Box 454002, Las Vegas, Las Vegas, NV 89154}
\ead{jason.steffen@unlv.edu}

\author{Jack J. Lissauer}
\address{Space Science \& Astrobiology Division, MS 245-3, NASA Ames Research Center, Moffett Field, CA 94035}

\begin{abstract}
We revisit the discovery and implications of the first candidate systems to contain multiple transiting exoplanets.  These systems were discovered using data from the \kepler\ space telescope. The initial paper, presenting five systems \citep{Steffen:2010}, was posted online at the time the project released the first catalog of \ik planet candidates. The first extensive analysis of the observed population of multis was presented in a follow-up paper published the following year \citep{Lissauer:2011a}.  Multiply-transiting systems allow us to answer a variety of important questions related to the formation and dynamical evolution of planetary systems.  These two papers addressed a wide array of topics including: the distribution of orbital period ratios, planet size ratios, system architectures, mean-motion resonance, orbital eccentricities, planet validation and confirmation, and the identification of different planet populations.  They set the stage for many subsequent, detailed studies by other groups.   Intensive studies of individual multiplanet systems provided some of \ikt's most important exoplanet discoveries.  As we examine the scientific impact of the first of these systems, we also present some history of the people and circumstances surrounding their discoveries.
\end{abstract}


\end{frontmatter}


\section{Introduction}

Among the early, groundbreaking discoveries of the \kepler\ mission were planetary systems where multiple planets are seen to transit.  Multi-transiting systems enable a wide variety of studies of the architectures and dynamics of planetary systems and the properties of individual planets within those systems.  Moreover, from 2012 until 2016, the majority of \ik planet discoveries (as opposed to simply planet candidates) were verified using unique aspects of multi-transiting systems, such as transit timing variations (TTVs) resulting from mutual planetary perturbations, or the intrinsically lower false positive probabilities when multiple planet candidates are present, to confirm or validate their planetary nature \citep{Lissauer:2012}.

The \ik spacecraft was launched in March of 2009, and science operations began two months later.  As of April 2009, only 36 multiplanet systems were known, 32 from radial velocity observations (see Figure 12.10 of de Pater \& Lissauer 2010), the 3-planet system orbiting the pulsar PSR B1257+12, the two-planet OGLE-06-109L system detected via microlensing, the directly-imaged HR 8799 system (with three planets known at that time, \citet{Marois:2008}), and the Solar System.  Most of the planets in these systems were more massive than Jupiter, and only seven, the pulsar system, five radial velocity (RV) systems, and the Solar System, had more than one planet less massive than Saturn.  Besides these seven, only a few systems had more than one planet with an orbital period less than one year, and most of the planets in the systems were widely-spaced---both in terms of physical space and orbital period ratio.

While we expected to see multi-transiting systems in the \kepler\ data, there was a collective sigh of relief when the first candidate multiplanet systems were observed.  While we anticipated that these systems would be interesting to study, their overall impact on the \kepler\ mission and on the field of exoplanets in general far exceeded our expectations.  The first announcement of multi-transiting systems \citep{Steffen:2010} (a.k.a. the ``Five Multis'') not only discussed these landmark systems, but it spawned several areas of subsequent study (e.g., orbital configurations, planet sizes, etc.)---giving some initial glimpses of the work to come.  The first statistical analysis of \kepler\ multis \citep{Lissauer:2011a} (a.k.a. ``Architectures I'', often abbreviated as ``Arch I'') presented watershed results in our developing understanding of the characteristics of planetary systems with orbital periods of $\lesssim 4$ months.  In this historical review, we recount early developments in the study of \kepler\ multis, focusing on the writing of these two key publications.

The Five Multis and Arch I papers presented some of the most important early results of the \kepler\ mission, and came at a time of both frenetic scientific activity and changes behind the scenes to team policies and organization.  The composition of the science team, the organization of the mission, internal communication and publication policies, follow-up observations, working groups, teleconferences, and scientific responsibilities were all being nailed in place when the \kepler\ floodgates opened.  Grappling with the enormous wall of flowing data was only one of the challenges faced by the science team.  Choosing the most important results, who should lead their publication, and who would contribute what was another.  At the time when the first major data release was fast approaching, the available window to address these issues and to produce results for the public and the exoplanet community was limited.  Regarding these two papers, reading through our emails of nearly a decade ago paints an interesting picture of how they came to be.

This manuscript both outlines the significance and legacy of these two papers and brings to light some of what was happening behind the scenes.  Section \ref{sec:ttv} reviews the history of transit timing variations studies prior to the launch of \ikt.  Most of the collaborative work in writing both Five Multis and Architectures I was done by members of the \ik TTV/Multi-Planet Working Group, the early history of which we summarize in Section \ref{sec:WG}.  In Section \ref{sec:fivemultis}, written largely in the first person, Jason Steffen presents his recollections of the conversations, emails, and historical events associated with the writing of Five Multis.  Section \ref{sec:multisovertime} reviews multis in subsequent planet candidate catalogs and provides an updated summary of our understanding of the five candidate multiplanet systems announced in the Five Multis paper. Selected scientific results from multis are discussed in Section \ref{sec:scifrommultis}.  Jack Lissauer summarizes events leading to Architectures I from his perspective in Section \ref{archhistory}.  Apart from Sections \ref{sec:fivemultis} and \ref{archhistory}, this review is written by both authors collectively.  Note that purely numerical dates of the quoted emails and elsewhere in this article are given in the US convention (MM/DD/YYYY).

\section{Transit Timing Variations Before \ik}\label{sec:ttv}

The possibility of using transit observables to measure the properties of planetary systems predates the works of \citet{Agol:2005} and \citet{Holman:2005}---the two papers generally credited with the development of transit timing variations.  There is little question that those papers marked seminal advances in the field.  But, as with any scientific milestone there was prior work.  For example, \citet{Miralda-Escude:2002} showed that transit duration variations due to precession could be used to infer the presence of an Earth-mass planet.  And, to quote William (Bill) Welsh, the study of ``eclipsing binary stars is ancient''.  Indeed, one paper that served as inspiration for \citet{Agol:2005} was \citet{Borkovits:2003}, which outlined the effects on the eclipse times from a hierarchical member of a triple star system.

One of the insights from the papers by \citet{Agol:2005} and \citet{Holman:2005} was that the direct terms in the disturbing function and mean-motion resonances can produce substantial TTV signals---large enough to probe for Earth-mass planets in systems with gas giants.  The first such analysis with this goal in mind  \citep{Steffen:2005} considered the Tres-1 system.  Within a few years of the primary TTV papers appearing, several studies of the TTV signal and its applications emerged.  Notable examples include using TTVs and transit duration variations (TDVs) to detect moons orbiting distant planets \citep{Kipping:2010}, Trojan planets \citep{Ford:2006}, precession \citep{Agol:2005,Heyl:2007}, along with a major advance in the analytic derivation of the signal \citep{Nesvorny:2008}.  However, despite the development of the theory of the TTV signal, and repeated attempts to find such a signal, no definitive detections had been seen by the time that these papers were written.  The lack of detections was primarily because most transiting planets known prior to the launch of \ik were hot Jupiters, which rarely have nearby perturbing companions \citep{Wright:2010,Steffen:2012b}, and because TTV systems tend to be in systems with smaller planets, and are therefore more difficult to find from instruments from the ground, or smaller-aperture space missions like ESA's CoRoT \citep{Auvergne:2009}.

A surprising fact is that the original motivation for the paper by \citet{Agol:2005} was to use TTVs, combined with measurements of the transit depth and Doppler amplitude, to determine the sizes of stars, rather than for the detection of planets or of characterizing planetary systems, and there remains a section of \citep{Agol:2005} devoted to stellar sizes.  The utility of TTVs for detection and characterization quickly became apparent.  At the same time, Jason Steffen's dissertation was supposed to be on using Very Long Baseline Interferometry to measure the shadow cast by the event horizon of the galactic center black hole on its accretion disk.  It was his great fortune that the ``practice problem'' of TTVs spilled beyond its original scope and developed into something valuable for a mission that was selected during his first year of graduate school, and that announced its Participating Scientist Program the year after he graduated.

\section{The \ik TTV/Multi-Planet Working Group}\label{sec:WG}

While we expected to find multi-transiting exoplanetary systems with \kepler , how many we would find, and what they would look like were unknown. The (rejected) 1998 \kepler\ mission proposal lists among its Expected Results for terrestrial planets ``70 cases (12\%) where $\geq 2$ planets per system are found''; however, this statement was not listed among the goals in the (selected) 2000 \kepler\ mission proposal \citep{Borucki:2016}.  Indeed, none of the six major goals of the \kepler\ mission (all of which were related to finding exoplanets and determining the properties of these planets and the stars that they orbit) mentions multi-transiting systems, although one goal was to ``Identify  additional  members  of  each  photometrically discovered planetary system using complementary techniques'' \citep{Borucki:2016}.  These complementary techniques were not specified, but presumably included RVs and likely also astrometry.  
It wasn't until after the \citet{Agol:2005} and \citet{Holman:2005} papers appeared that the large variations calculated from the disturbing function demonstrated the viability of TTVs for \kepler .


In 2007, the \ik mission solicited proposals to join the mission via the Participating Scientist Program.  Of the selected proposals, two were directly related to analyzing the TTV signal (Jason Steffen and Matt Holman), and two more were tangentially related to TTVs (William Welsh on estimating transit times and Eric Ford on lightcurve analysis and the eccentricity distribution).  For TTV proposals specifically, there were two primary issues that they addressed.  The first was measuring planet masses that were too small to detect with Doppler spectroscopy.  The second was detecting non-transiting planets due to TTVs induced on their transiting counterparts.  While eventually both of these items would be addressed with \kepler\ data, we realized that systems with multiple transiting planets were far more rich in information than single or isolated planets that showed TTVs.  The transit signals from the multiple planets broke a variety of model degeneracies (typically due to different resonances) that plague single-planet systems.


At the  \kepler\ Science Team Meeting (STM) in November 2007, shortly after the selection of the Participating Scientists, or ``PSPs'' as they became known, we recognized our overlapping interests and the benefit of working together to accomplish the goals of our respective proposals.  While several of the future members of our group weren't able to attend this early meeting, two months after it took place, Jack sent an email to Jason as well as Eric Ford, Laurance Doyle and Matt Holman:
\medskip

\noindent{\bf 1/24/2008: Email from JJL:}
\\
{\tt \footnotesize 
Subject:        Kepler multiple object systems and dynamics investigations\\
...\\
\noindent
Welcome to the Kepler Science team!\\  

\noindent
Our Kepler duties are somewhat related, so we should keep in touch and coordinate our work when appropriate...\\

\noindent
We should have a breakout session at the next Kepler Science team meeting where most or all of us are in attendance...}
\medskip

We scheduled a time to discuss our shared interests for the science team meeting that was to occur that coming May---juxtaposed with the IAU symposium on exoplanets (IAU Symposium 253 ``Transiting Planets'', held in Cambridge, MA, May 19-23, 2008).  Thus began the long and fruitful collaboration of what became the \ik Multibody/TTV working group.

Our second meeting was for a few hours on the afternoon of Veteran's Day 2008, the day prior to the start of an STM at Ames.  However, it was our third meeting, in March 2009, where we got down to the brass tacks of collaboration.  This meeting was coincident with the \kepler\ launch and the key discussion, surrounding a dinner table, was Eric Ford, Jack Lissauer, Bill Welsh, Matt Holman, and Jason Steffen.  There we hashed out details regarding who needed what information to do what science, the scope of our first projects (i.e., where everyone's toes were located so they could be avoided), and how we would share our results with each other.

Of course, some of this discussion was wishful thinking.  The team had been so focused on preparing for launch that the data sharing policies hadn't been updated.  They did not yet reflect the change that comes when transitioning from a proposed project competing for resources to an operating mission.  Data access was still on a need-to-know basis with tight controls.  For example, here is a portion of an email regarding access and sharing that was written more than four months after launch:

\medskip

\noindent{\bf 7/21/2009: Email from JHS:}
\\
{\tt \footnotesize 
Subject:        Data Request -- for TTV analysis\\
Date:   Tue, 21 Jul 2009 12:11:25 -0500\\
From:   Jason Steffen\\
To:     Borucki, William J., Batalha, Natalie, Gautier, Thomas N\\
CC:     Welsh, William F., Lissauer, Jack J.\\

\noindent
Hi Bill, Natalie, and Nick,

I'd like to make a request for some Kepler data for work that I will do in coordination with William Welsh.  



\centerline{$\vdots$}

\noindent
PS. Do I need permission to share these data with Matt Holman?
}

\medskip

As one can see from the postscript, even communication within the team was tightly regulated.  Eventually we sorted out the \kepler\ data sharing issues.  Our group expanded to include collaborators, postdocs, students (most notably Dan Fabrycky, Darin Ragozzine, and Jerry Orosz), as well as other members of the \kepler\ science team who were interested in our work (Dave Latham, Dimitar Sasselov, Geoff Marcy, and Bill Cochran).  Many of these additional members were invited to join the group in an email dated April 21, 2010---right in the middle of the developments we are presenting here:

\medskip

\noindent{\bf 4/21/2010: Email from Eric Ford:}
\\
{\tt \footnotesize 
Subject:        Multiple System \& TTV Working group\\
Date:   Wed, 21 Apr 2010 19:02:45 -0500\\
From:   Eric Ford\\
To:     Matthew Holman, Jason Steffen, Jack Lissauer, William Welsh, Darin Ragozzine, Althea Moorhead, David Latham, Jason Rowe, Ronald Gilliland, Geoff Marcy\\

\noindent
Dear Kepler folks with an interest in multiple planet systems and/or TTVs,

An unofficial working group (Holman, Ford, Lissauer, Moorehead,
Ragozine, Steffen, Welsh) has been holding a teleconference most
Wednesdays Noon ET (9am PT) since January.  Discussions have included
both multiple planet systems and TTV issues.  At today's SWG telecon,
we were encouraged to form more formal working groups, including two
particularly relevant to us on:
- Multiple Systems, and
- Transit Timing for Detection of Exoplanets

This raises a several questions:

\noindent
1.  Are people generally supportive of bifurcating into these two
working groups?  Should it be more than two?  [I agree that two is
probably a good idea, but fear three will result in too many
telecons.]

\noindent
2.  Who wants to participate in which group(s)?  [I will try to
participate in both.]

\noindent
3.  Are there other people within the science team (or their
associates) who we should ask to join us?  [My understanding is that
are telecons will be advertised and open to the fuull science team and
their associates.  As far as I can tell, inviting them now only means
they get to influence the time of telecons and perhaps the choice of
the chair.]

\noindent
4.  What times can work for each of these working groups?  [I suggest
one group take over the Wed noon-1pm ET slot and the other group look
for a time on a Mon or Fri.  I can setup a doodle poll, if people
would like.]

\noindent
5.  Who should chair each of the working groups?

Hopefully, we can agree to a plan via email before April 28.  If not,
then I suggest that we have another combined telecon at the old time
of noon ET (9am PT) on Wed, April 28.  One of the agenda items can be
resolving the above issues.

Thanks,
Eric
}

\medskip

Shortly thereafter we divided our working group into two---one focusing on TTVs specifically and the other on the properties of multiplanet systems.  Jack chaired the multibody working group and Jason chaired the TTV working group (after a brief discussion with Matt Holman).  As time went on, this division was mostly on paper as the member lists were virtually identical, and we shared the same email listserve, ``kepler-ttv'' (eventually, the two groups officially merged into one).  If nothing else, this division served as a means to keep our weekly---or twice weekly---telecons moving with fresh topics for discussion from the different scientific perspectives.

Given this backdrop, we established our group and our culture of working together.  Beginning with the two papers discussed in this manuscript and the discovery papers of Kepler-9 and Kepler-11 (whose histories are recounted in \citet{Ragozzine:2019} and \citet{Fabrycky:2019}, respectively), we made progress on a variety of fronts.  By the time of this writing, our broader group would produce some three dozen papers with over 5000 citations.  Even now the majority of the initial working group continues to collaborate on the analysis of \ik data.  

In addition to research, the \ik TTV group was tasked by the  mission to select targets to downlink Short Cadence (SC) data (summed over intervals of one minute rather than the 30 minutes of typical \ik Long Cadence data) to allow for more accurate measurements of transit times.  A very limited number of SC target slots were available for this purpose, so no SC data were used for statistical studies of the type discussed herein.  However, SC data were very useful in improving the accuracy of TTV measurements and planetary mass determinations therefrom \citep[e.g.,][]{Jontof-Hutter:2016}.

\section{The First Multi-transiting Systems}\label{sec:fivemultis}

\noindent
{\it This section is written from the viewpoint of JHS}

Our first paper on multi-transiting systems came after a period of frustration in the community and considerable anticipation for results from \kepler .  Two years before \kepler 's launch, an exoplanet workshop in Heidelberg (September 2006) portended major advances for planets, but ended up being a disappointing, week-long discussion of red noise in transit surveys.  

Nevertheless, those issues were addressed and the field continued to advance.  Indeed, only a year and a half later, at the IAU Symposium No. 253 mentioned above, Michel Mayor announced his finding that 30\% of F, G, and K stars had a small planet within 50 days.  Jack told me that he thought it was the most important result of the conference---a conference with sufficient land-mark results that several people referred to it as the Woodstock of transiting exoplanet science (with the organizers dressing in Woodstock-themed costumes at the close of the conference).  My perception of this Symposium was that it was a major release of pent-up frustration from the exoplanet community.  As I saw it, the Boston conference, with all its trappings, was what we expected to have had back in Heidelberg.

While the discoveries continued to mount, theoretical models of dynamical evolution of hot Jupiter systems suggested that the capture into Mean Motion Resonance (MMR) of residual planetesimals would often produce terrestrial planets both interior and exterior to the Jupiter \citep{Zhou:2005,Thommes:2005}.  At the same time, results from RV surveys were showing a high frequency of sub-Neptune planets with short orbital periods of several tens of days.  Given this situation, when the \kepler\ data showed several systems with multiple transiting candidates, it was a gratifying reassurance that we weren't completely misguided.




It took nearly a year from the time of launch for the science team to sort out the internal lines of communication with the new Participating Scientists---getting people onto the right telecons, looking at the right data, and sharing the right documents.  The first time that multi-transiting systems rose above the noise was in early April 2010, about 10 weeks prior to the deadline when the first data release would be made public.  At that time, I requested that multi-transiting systems be included in the agenda for an upcoming team meeting.  Even though the scientific value of multi-transiting systems was recognized, the sheer volume of essential labor pushed a paper announcing the discovery of multi-transiting systems into the background.  Between vetting the planet candidates, doing follow-up observations for stellar multiplicity, contamination, stellar properties, RV mass measurements, statistical noise, and telecons there was little time for more mundane tasks like authoring.


Further driving the multi-transiting systems onto the back burner was a desire to showcase some of our more exciting individual candidate planets and planetary systems.  Everyone wanted to produce spectacular results since it was more than one year after launch and, at that time, the published exoplanet discoveries by the \kepler\ mission were only the five planets that had been announced at the AAS meeting in January 2010---four hot Jupiters and one hot Neptune.

In April 2010 there was an exoplanet conference held in Obergurgl, Austria.  It was clear at that conference that people were getting restless.  I shared, in an email, the sentiment that there was a ``palpable let-down in Austria due to the continued silence from Kepler'' to which others who attended agreed.  There were even discussions, conveyed to us through the grapevine, of people who were threatening to boycott talks by \kepler\ scientists.  We sensed, and felt, an urgent need to produce material that would be worth the wait.

Among the \kepler\ target systems that were being developed at this point in the calendar was the first ``heart-beat star'', KOI-54, which was initially modeled as a black hole/stellar binary.  KOI-126, a triple star we initially thought was a double planet.  And, the first system showing Transit Timing Variations (TTVs).  Both Matt Holman and I were brought to the science team to conduct TTV studies for the mission.  We were both keen to lead the first TTV analysis of an obviously real signal---KOI-377.  We both knew what it would mean to each other.  However, there wasn't a turf war or some kind of competition.  Rather, to my mind, we wanted to resolve the issue in a way that would preserve the good feelings in the group and would be beneficial to both.

In a phone call between Matt and me on April 23, 2010 we agreed that Matt would lead the first TTV analysis of that system (the future Kepler-9) and I would later lead a paper on the first non-transiting planet discovered with TTVs.  (Ultimately, other circumstances prevented this side of the agreement from being realized.)  For full disclosure, at the time I thought that KOI-103 as the most promising candidate for a comprehensive analysis with KOI-142 and KOI 646 as other possibilities.  Eventually, KOI-142 did see the light of day as the Kepler-88 system \citep{Nesvorny:2013}; see also the history of the discovery of  that system in this issue \citep{Nesvorny:2019}, KOI-646 turned out to be a triple star system, and KOI-103 remains an unverified planet candidate.




With the issue of Kepler-9 addressed, with our return from the conference in Obergurgl, with the organization of the science team settled, and with the calendar still moving toward the date for the data release, I wanted to contribute something meaningful to the mission and, therefore, turned my attention to the hind-most burner where the multi-transiting systems were simmering.  
On May 21, 2010 (24 days to submission) Jason Rowe provided the first set of candidates for us to consider including.  This list comprised KOIs 137, 152, 157, 191, 209, 686, 877, 896, and 941.  After a few days, and about a hundred individual emails, the list was reduced to KOIs 152, 191, 209, 877, and 896.  Two notable systems that were removed for further scrutiny eventually appeared as Kepler-18 \citep[KOI-137][]{Cochran:2011} and Kepler-11 \citep[KOI-157][]{Lissauer:2011a}.  At the time of the Five Systems paper, the future Kepler-11 only showed four planet candidates rather than the six that were announced six months later.  The remaining targets were selected in part to show the variety of planet sizes, orbital configurations (especially pairs near mean-motion resonances), and differences in the number of observed planet candidates.







Given that we now had a sample of candidate multi-transiting systems to announce, we still had to decide what to do with them.  The content of the Five Systems paper evolved constantly throughout the authoring process.  One item on everyone's mind was TTVs.  However, a simple prediction using a Monte Carlo simulation of possible TTV signals alone was not a suitable result to accompany such an important discovery.  Nevertheless, the section on TTVs did serve as the kernel from which the rest of the manuscript grew.  As various members of the science team considered what they had to offer, different sections of the paper began to appear.  This kludged effort, as we all attempted to grapple with both how to study multi-transiting systems and how to present the related findings, set the stage for much of the subsequent literature---including the Architectures I paper that we will address later in this work.

Three weeks before the submission deadline the observers in the Kepler Follow-up Observation Program (KFOP) began completing the reconnaissance observations of the systems---taking spectra for stellar classification and seeing-limited images to identify contaminating background stars.  This work was accomplished by Bill Cochran, Geoff Marcy, and Dave Latham.  At the same time, Eric Ford provided me with an estimate of the eccentricity distribution from RV planets that I used to conduct the TTV Monte Carlo simulation.  An important piece of information that I needed for this simulation was constraints on the planet masses, which were initially provided, two days later, by Jonathan Fortney with input from Dimitar Sasselov.  The final mass estimates came after what seemed like a long time (it was only 10 days later---but that itself was only 10 days before the deadline).  While that process got started, more people volunteered to contribute different sections to the Five Systems paper.  Darin Ragozzine offered to ``calculate the probability that the outer planet transits given that the inner planet does as a function of mutual inclination''.  And, Fran\c cois Fressin began a BLENDER analysis---an analysis of the lightcurve and stellar properties used to determine the probability that a transit signal is an astrophysical false positive. (See the paper in this issue by \citealt{Torres:2019} for additional information on BLENDER.)

Initially, the plan was to announce Kepler-9 at the same time as the data release catalog and the Five Multis paper.  However, with just two weeks remaining it became clear that Kepler-9 wouldn't be ready---especially if NASA planned to have a press release to accompany its announcement, which they eventually did.  Even with this delay in the release of Kepler-9, given the excitement surrounding the observation of TTVs, the bulk of the effort in our group was still devoted to Kepler-9 and would be almost up to the time when the Five Multis paper was submitted.


Unrelated to \kepler , but still filling the same calendar, was the fact that I was active in particle cosmology research at Fermilab and was leading a laboratory test of dark energy \citep{Steffen:2010b}.  Two weeks before the submission deadline I attended a week-long dark matter conference in Leon Mexico (Dark Side of the Universe, June 1-6, 2010).  I began the final TTV simulations prior to departing, and a lot of my time at the conference was devoted to finishing the Five Multis paper.

One crucial email from Eric Ford arrived during that conference on June 2.  He shared work from his student Robert Morehead on the development and first application of the $\xi$ statistic (the email used $\phi$).  This statistic is defined as 
\begin{equation}
\xi \equiv \frac{D_\text{in}}{D_\text{out}} \left( \frac{P_\text{out}}{P_\text{in}} \right)^{1/3},
\end{equation}\label{eq:xi}
\noindent
where $D$ is the transit duration and $P$ is the planetary orbital period (with the subscripts ``in'' and ``out'' denoting the inner and outer planet in a given pair). The $\xi$ statistic, which eventually was employed to measure properties of the distribution of orbital eccentricities \citep{Lissauer:2011a,Fabrycky:2014}, was used in the Five Multis paper to bolster the claim that these planet candidates were orbiting the same star.  In the email Eric stated ``If the two candidates were orbiting stars of significantly different densities, then this ratio could significantly deviate from the expected range.''

As the dark matter conference wore on, and the deadline approached, results started to arrive at what seemed an agonizingly slow pace, but was actually rapid succession.  On Thursday June 3 Andrew Howard and Geoff Marcy provided their reconnaissance spectra at 9am, light curves and transit times came from Dan Fabrycky at 3:30pm, and text for the planet properties by Dimitar Sasselov at 10pm.  Noon the next day (June 4) brought observations from Steve Howell and introductory text from Jack Lissauer.  Stellar properties came from Geoff Marcy at 7:30pm, and at 9pm Darin Ragozzine sent his coplanarity analysis and Natalie Batalha sent the initial Data Validation results from the Kepler pipeline.  (All times here are Central Daylight Time.)

An unfortunate dinner on Friday night forced me to spend the next day (June 5) near the facilities of my hotel room.  That day became the most productive single day in the authoring process, where I incorporated the information I had received, produced a more complete draft, and shortened the list of remaining tasks to only a handful of items.  The final $\xi$ analysis, final text on planet properties, and final Data Validation analysis arrived on June 6 as the work was wrapping up.  On the one day between the Dark Matter conference in Mexico and the Science Team meeting in Denmark, I will neither confirm nor deny that I called my graduate advisor (Eric Agol) to tell him what I was working on.

Just past midnight, at 12:30am on Monday June 7 (in coincidental celebration of the 66th anniversary of the liberation of Bayeux following the Normandy landings)
, the draft paper was submitted to the \ik Science Council for review.  At this time in the \kepler\ mission, the Science Council was tasked with reviewing all papers that were to be submitted by the broader science team.  While the Five Multis paper wasn't yet complete, it was close---and the remaining details would have to wait until the Science Team met in Aarhus, Denmark the next day (one week before the data and the paper would go public).

The final week brought in the last of the needed results as well as its own share of surprises.  On June 9, Eric Ford found a mission status report from NASA that leaked to the public that the mission had discovered several multi-transiting systems---a deflating situation that, fortunately, didn't garner much attention.  Fran\c cois Fressin gave an update on the BLENDER analysis on June 8 with the final results arriving from Willie Torres on June 11 (three days before submission).  As a final twist, at the beginning of our week-long meeting, Jason Rowe---who had been worked to the bone preparing for the data release, who was a central figure in preparing the larger catalog, and who was providing us with essential text about modeling the light curves---seemed to have disappeared from the planet.  Everyone knew he had arrived in Denmark, and yet he was no where to be found.  Many members of the science team were asking about him---especially those from NASA Ames, who knew how crucial his work was to the mission.  Without any responses to our emails or phone calls, there was serious concern for his whereabouts and health.  On June 12 he emerged at 4am from a 36-hour nap.  Back to his prolific self, he provided his figures and text on the five systems in short order.  Twenty hours later, at on June 14 at 16:42 CDT the paper went to arXiv.  

At the same time, the paper was submitted for publication in the {\it Astrophysical Journal} where it made its way through the review process, was accepted in October of 2010 and appeared online in November of the same year.  For more than two months, following its appearance on the arXiv, the Five Systems paper was the most highly read paper on the NASA Astrophysics Data System, and it remained highly read until it was superseded by the more comprehensive analysis of the Architectures I paper.

\section{Kepler Multi-Transiting Systems Over Time}\label{sec:multisovertime}









\subsection{Planet Candidate Catalogs}

With each new catalog of planet candidates released by the \kepler\ mission, the number and nature of multi-transiting systems evolved.  Although the number of multi-transiting systems has increased over time, changes in the vetting procedures has caused some systems to creep in and out of the planet candidate list with some candidates no longer showing up as Threshold Crossing Events.  A prime example of a missed system is Kepler-9, which needed to be put back on the candidate list by hand with each new catalog.  (See \citet{Ragozzine:2019} for a history of Kepler-9.)

One reason for these changes in the classification of many systems (including Kepler-9) is the presence of TTVs that distort the shape of the transit when the data are folded on a constant orbital period.  The mismatched ingress and egress for the various transits makes the event in the folded data file look more like the canonical V-shaped eclipse of a binary star.  The effect often became more important over time because the curvature in the TTV signal can take several quarters of data to be visible.


\subsection{Current State of the Five Systems}

Presently, all of the initial candidate systems in the Five Systems paper have at least two of their planets confirmed or validated.  Four of the five systems have additional planets beyond those seen in the first two quarters of \kepler\ data (upon which the Five Systems paper was based).  The planets in these systems were confirmed using a variety of methods---both dynamical and statistical.  For example, two of the systems were confirmed using TTVs in the study by \citet{Xie:2013}---following the methods of \citet{Steffen:2012b,Fabrycky:2012a,Ford:2012a}, and \citet{Steffen:2013a}.  One system is the three-planet, KOI-877 (Kepler-81) system.  The other is KOI-152 (Kepler-79), which initially had only three known candidates, but is now a four-planet system where all planets are confirmed.

Two of the five systems were confirmed using the statistical properties of multi-planet systems.  The expected higher reliability of planet candidates in multis was discussed in both \citet{Latham:2011} and the Architectures I paper.  \citet{Lissauer:2012} quantified the increase in reliability of multis and used it as well as the apparent flatness of the 5-candidate KOI-707 system to validate these candidates as the 5-planet, Kepler-33 system.  The method used the fact that false positives are uncommon, and unlike real transiting planets, they are not expected to cluster in systems that have other false positives or planet candidates.  Multiple false positives in a single system is rare, and  multi-transiting systems are unlikely to be false positives.  This method for validation by multiplicity was further fleshed out in \citet{Lissauer:2014} and used in the companion paper by \cite{Rowe:2014} to validate the two-planet KOI-209 (Kepler-117) and the three-planet KOI 896 (Kepler-248, though only two of the three are presently validated).

Finally, the KOI-191 (Kepler-487) system has four planet candidates. Two of these candidates  were validated (KOI-191.01 as planet ``b'' and 191.04 as planet ``c'') by showing that the planet hypothesis is much more probable than the likelihood of being false positives caused by eclipsing binary stars using the VESPA code \citep[][which considers only astrophysical false positives and therefore can be somewhat less reliable than the validations performed by \citet{Rowe:2014}]{Morton:2016}.  The KOI-191 system remains an interesting case study as it has a large gas giant embedded in a system of smaller planets---similar to the WASP-47 system \citep{Becker:2015}, even including an ultra short-period planet on a 17-hour orbit \citep{Sanchis-Ojeda:2013,Steffen:2015}.

\section{Science from Multi-transiting Systems}\label{sec:scifrommultis}


The Five Systems paper addressed several topics related to the properties of planetary systems and their detection and characterization.  The analysis in that paper was expanded in later studies and applied to more complete catalogs of \kepler\ planetary systems.  In this section we look at some of the first forays into understanding the nature of multiplanet systems that were presented in the Five Systems paper.


\subsection{Period Ratio Distribution}

A straightforward observable for multi-planet systems is the ratio of planetary orbital periods.  When moons or planets undergo convergent migration with respect to each other, through tidal interactions with the primary or with a gas or planetesimal disk, they can be captured into Mean-Motion Resonance, or MMR \citep{Goldreich:1965,Peale:1976,Lee:2002}.  Thus, the presence or absence of planet pairs near MMR gives insights into the dynamical history of the system.  The period ratios for the systems in the Five Systems Paper were included for this reason.  A particularly valuable aspect of period ratios from transiting systems is that the orbital period for transiting systems can be measured with a precision that is two or more orders-of-magnitude better than by other means (e.g., RV).  Consequently, planet-planet interactions are visible in a wider range of systems than what can be observed with other techniques.

For the five systems, at least three planet pairs were close to MMR (two near 2:1 and one near 5:2) with the others being quite far from any resonance.  The hand-picked systems were not representative of the whole population, but they were chosen in part to showcase the variety of period ratios seen in the \kepler\ data at the time.  When the first full catalog of planet candidates was released 7 1/2 months later, we expanded upon the results from the Five Systems paper in Architectures I where we made our first attempt at analyzing the observed period ratio distribution \citep{Lissauer:2011a}---with a follow-up analysis in the  \citet{Fabrycky:2014}.  Eventually, \citet{Steffen:2015} calculated the period ratio distribution after correcting for the reduced probability of detecting more widely separated planet pairs due both to geometry and pipeline completeness.

Several other features appear in this distribution.  One example is the overabundance of planet pairs near a period ratio of 2.2 \citep{Steffen:2015}.  Another is the interesting population of isolated planets with short orbital periods that may be related to the large population of single planets with short orbital periods \citep{Sanchis-Ojeda:2013,Lissauer:2014,Steffen:2016}.  Finally, we find that planet pairs become more widely separated when the innermost planet has an orbit of a few days or less \citep{Steffen:2013c}.  This last feature may be explained by tidal interactions within the system \citet{Lee:2017} or secular chaos \citet{Petrovich:2018}.

To study the properties of the period ratio distribution near resonance in the Arch I paper, we developed a quantity $\zeta$ that stretched the intervals between all MMRs of a given order so that they span the interval $(-1,1)$.  The general form was later published in the second paper in the series \citep[Architectures II,][]{Fabrycky:2014}.  This quantity allowed us to effectively stack all first or second-order MMRs together to look for common features (though it admittedly may not have an important physical interpretation).  Several other quantities for measuring the distance from resonance are outlined and discussed in \citet{Steffen:2015}.  

\subsection{Eccentricity Distribution}

Another quantity introduced in the Five Systems Paper is the normalized transit duration ratios in multi-planet systems, embodied in the $\xi$-statistic (Equation \ref{eq:xi}).  This quantity can help eliminate some false positive scenarios since a ratio far from unity for a pair of planet candidates implies that they orbit host stars of different densities.  A second, and more widely used, application is that $\xi$ can constrain the distribution of orbital eccentricities and inclinations.

Both eccentric and inclined orbits will affect the observed duration of planetary transits.  For eccentric orbits, the changing speed of the planet as it passes from pericenter to apocenter and back changes the transit duration.  If the planets in a system have large, randomly oriented eccentricities, then the duration ratio distribution will spread out.  At the same time, mutually inclined orbits will have planets that transit across different chords of the stellar disk, which produces different transit durations.  \citet{Fabrycky:2014} used the distribution of the $\xi$ statistic to constrain both the inclination and eccentricity distributions and estimated typical eccentricities of 0.03 and inclinations of 1.6$^\circ$ for \kepler\ planets.


\subsection{Confirmation and Validation}

False positives, especially from background eclipsing binary stars, is one of the challenges faced by transiting planet surveys.  Eliminating astrophysical false positives often requires a variety of ground-based follow-up observations---an intensive regimen that is not economical for the large number of (usually dim) systems identified by \kepler .  During the first years of the \kepler\ mission, we wanted to produce bona fide exoplanets rather than simply planet candidates.  Claiming that a candidate is actually a planet can be done either by confirmation, which we collectively defined as using dynamical observables such as Doppler or TTV measurements to verify the planet nature of the candidate, or by validation, which uses only statistical arguments to show that the observed transits were far more likely to be caused by a planet than by anything else.  Because of the desire for high confidence in our candidate detections, several scientists developed statistical techniques to eliminate false positives.  Among the first of these techniques was the BLENDER analysis, which was initially used to validate the planet nature of Kepler-9d \citep{Torres:2011}.  A preliminary version of the BLENDER code was applied to the Five Systems.  That analysis, which would eventually play a significant role in many \kepler\ discoveries \citep[more on this history is in][]{Torres:2019}, was provided by Fran\c cois Fressin and Guillermo Torres.

Multi-transiting systems also enabled the dynamical confirmation of planets through TTVs.  Specifically, the TTVs between planets in a system are generally anti-correlated, and the statistical significance of the anti-correlated TTVs can demonstrate that transiting planet pairs dynamically interact, implying that they are in the same system.  Once established, requiring dynamical stability within that system can constrain the masses of the objects to be within the planetary regime.  A series of papers introduced this confirmation procedure \citep{Ford:2012a,Steffen:2012b,Fabrycky:2012a} and a subsequent analyses added to the planet tally \citep{Steffen:2013a,Xie:2013}.  For a few years this method confirmed more \kepler\ planets than any other technique.

At the same time that the TTV confirmation method was developed, the procedure to validate planets based upon planet multiplicity was devised by \citet{Lissauer:2012}.  Here, while an individual planet candidate might have a modest probability of being a false positive, the probability declines significantly if multiple planet candidates are seen on a single target---implying that planet candidates in multiplanet systems are unlikely to be false positives.  Multiplicity arguments eventually validated several hundred planet candidates in two significant papers \citep{Lissauer:2014,Rowe:2014}.

\subsection{Transit Timing Variations}

The Five Systems paper had a section on expected TTV signals from the various systems---using our best estimates of the planet properties.  This particular section was the first one conceived and served as the seed from which the expanded scope eventually grew.  The development of TTVs, the manifestation of planet-planet interactions within the system, as a tool to understand planetary systems had several early motivations.  Indeed the initial motivation for the study by \citet{Agol:2005} was to use TTVs as a means to measure the stellar sizes.  Only after working on the problem for a while did mass measurements, dynamics, and planet detection come to the forefront.

TTV applications evolved considerably with the arrival of \kepler\ data.  Initially, much of the motivation to study TTVs was to detect non-transiting planets.  However, as the data rolled in, and the number of multiplanet systems with TTVs grew, the TTV studies shifted toward characterizing the planets in multi-transiting systems.  For example, as indicated above, while the Five Systems paper was being written, our working group actively pursued an analysis of the TTVs of Kepler-9---the first planets with a definitive TTV signal.  The change to planet characterization over planet detection was motivated by the unambiguous nature of the TTV signal in multiplanet systems versus the signal in a single-planet system.  Specifically, knowledge of the orbital period and phase of the perturbing planet significantly reduces the difficulty in interpreting the TTV signal.  Two other papers in this issue discuss the history of the first TTV analyses with \kepler\ data for both single planet, \citep{Nesvorny:2019} and multiplanet systems \citep{Ragozzine:2019}.

Eventually, the pressing need for transit times to analyze led to the production of a series of TTV catalogs---or tables of transit times for analysis.  This important, and often unsung, work was done primarily by two parties---Jason Rowe on one side and Tsevi Mazeh and his group on the other (notably Tomer Holczer).  Many early analyses, including the planet confirmation papers listed above, used the Rowe transit times, which were generally internal data products due to the lack of time to write a publishable catalog.  Most later studies used the catalogs from \citet{Mazeh:2013} and \citet{Holczer:2016}.  Ultimately, the analysis of many systems used neither catalog and instead modeled the lightcurve directly with a ``photodynamical'' model---a model that uses the dynamics of the system to calculate the light curve at every point in time instead of just during the transits \citep[e.g.][]{Carter:2012}.

\subsection{Other Science}

A number of other investigations were enabled by multi-transiting systems.  We give two notable examples here that were explored in subsequent papers from the \kepler\ science team.  One is the relative sizes of planets in a system.  With all of the planets in a multi-transiting system orbiting the same star, the ratios of the planet sizes are less prone to systematic errors.  KOI-191 (now Kepler-487) was chosen for the Five Systems paper specifically because of its unusual ratio of planet sizes---having a large Jovian planet just outside a smaller inner one.  Today, the distribution of planet size ratios indicates that the majority of planets are similar in size one to another with relatively few exceptions.  In general, the outer planets are slightly larger than their inner counterparts---though not by much \citep{Ciardi:2013,Weiss:2018}.

The second example is examining the differences between single transiting systems and multi-transiting systems, since there may be different populations of system architecture that point to different dynamical or formation histories.  Direct comparisons of the singles and multis began with \citep{Latham:2011} which showed that small 2--4$R_\oplus$ planets are the most common types of planets in all \kepler\ systems and that the presence of gas giants largely precluded the presence of smaller planets (see also \citet{Steffen:2012b}).  Subsequent studies examined the differences in the distribution of planetary eccentricities \citep{Moorhead:2011} and orbital periods \citep{Lissauer:2014, Steffen:2016}.  Ultimately the various system architectures, including multiplicity, were examined in detail with the ``Architectures'' papers---the first of which we discuss next.

\section{Architectures I}\label{archhistory}

\noindent
{\it This section is written from the viewpoint of JJL}

\subsection{Planning} 
By the time of the \ik Science Team Meeting in \AA rhus, Denmark during June 2010, the Science Team (in this case, primarily Jason Rowe), had already identified several dozen candidate multi-planet systems.  Orbital periods were measured very precisely, and size estimates (some quite uncertain) were available.  It was clear that multis tended to lack the giant planets commonly identified in singles.  As a planetary dynamicist, I was particularly excited by the sheer numbers of multis detected, as well as the precision at which the planets' orbital periods were known. 

During the \AA rhus meeting, both Dave Latham and I expressed interest in leading team papers focusing on statistical analyses of the \ik multis. Dave was primarily interested in the differences in properties of the planets themselves (the paucity of large planets and lack of hot jupiters among the multis), while my interests were focused on the relationships among planets in the same system. Therefore, we decided to divide up the science and write two complementary papers, both of which were published in 2011.

By the end of September 2010, the research for the paper that we now refer to as Architectures I was already taking shape, as was the plan for who would do which tasks. The research was being organized by the \ik TTV/Multi-Planet Working Group. An excerpt of the first detailed message I sent to this group on this project shows how far we had progressed in our planning:
\smallskip

\noindent{\bf 9/29/10: Email from JJL to the team:}
\\
{\tt \footnotesize 
Subject:     [kepler-ttv] outlines of koi-157 and multi-planet statistics papers\\
Date:     Wed, 29 Sep 2010 20:22:00 -0500\\
From:     Jack Lissauer\\
To:     Daniel Fabrycky \\
CC:     kepler-ttv\\
...\\
Multi-Planet Statistics Paper Outline\\
1)    Introduction (JL)\\
2)    Lightcurves (JR)\\
a.    TTVs\\
3)    Parameters\\
a.    Stars\\
b.    Planet candidates (JR)\\
4)    Confidence level (inc. validation of some?) JL\\
a.    Randomization of periods tests (DF)\\
b.    Stability tests\\
i.    Analytic (2 planets w/periods) JL\\
ii.   Numerical (3 planets w/periods) DF\\
5)    Inclination distribution (DR \& JL)\\
6)    Conclusions (JL)\\
}

\noindent (The portion of text excised from the above email appears in the companion review of the discovery of the Kepler-11 system \citet{Fabrycky:2019}.)  Most of the topics listed  in this message were covered in Arch I, and all of the people listed contributed to the paper in a major way.

\subsection{Research and Writing the First Posted Version}
As with the Five Multis paper discussed in Section \ref{sec:fivemultis}, Architectures I was produced under severe time constraints.  It was part of a group of \ik papers that were posted to arXiv.org on Wed, 2 Feb 2011 in the hour prior to the deadline to go live that evening, less than 24 hours after the first data release to include all \ik targets and the same day as the NASA press conference that announced the discovery of the Kepler-11 system \citep{Lissauer:2011b} and the planet candidate catalog associated with the data release \citep{Borucki:2011}.  

\citet{Latham:2011} and Arch I both concluded that multis are significantly more likely than singles to represent real planets independently and using different arguments. But neither study highlighted this result, and the primary authors didn't realize that the other paper included it until well after both papers were published despite being coauthors on each others' papers! \footnote{The \ik team was publishing a vast amount of material from 2010 -- 2013, reaping the benefits of more than a decade of hard work prior to launch. Thirteen papers reporting different scientific results from \ik  with both Dave Latham and I included in the author list were published in refereed journals in 2011 alone.  The primary contributor to a paper (or to the acquisition or analysis of the data used therein) was generally listed as lead author, with the other major contributors listed in decreasing order of contribution, followed by one or more alphabetical lists of people who made lesser (in many cases exclusively indirect) contributions to the research.  Dave Latham and I were placed near the end of the author lists of the paper that the other of us led; I emailed Dave a few comments on a draft version of the \citet{Latham:2011} paper, but don't have any records of having made a substantial direct contribution.}

I worked the longest hours in my life during the period leading up to the  1 Feb 2011 \ik data release, and indeed extending a few weeks beyond it. I was leading two major \ik team research projects that resulted in landmark papers and was an active participant in several others.  The Kepler-11 discovery paper, which I had spearheaded from the beginning, absorbed most of my time until it was accepted by {\it Nature} on 20 Dec 2010; development of the art and preparation for the press activities associated with that paper required a significant amount of time from mid-December through early February \citep[see more information in the history][]{Fabrycky:2019}.  I took only one day off between Thanksgiving (2010 November 25) and sometime in the latter half of February 2011. On New Year's Day, I spent $\sim 5$ hours on a general \ik Science Team telecon focused on the \cite{Borucki:2011} planet catalog paper (the telecon lasted a total of $\sim 6$ hours, but I missed almost an hour because I went on a hike during the break and was gone longer than expected). Despite the demands of the Kepler-11 paper and its associated publicity on my time and the time of Dan Fabrycky (the third author of Arch I), by late January much of the research for Architectures I had been completed and we were busy assembling our results into a manuscript. 

\smallskip
\noindent{\bf 1/23/11, 9:02 AM: Email from JJL to primary co-author:}
\\
{\tt \footnotesize 
Subject:     football to be passed back in < 3 hrs\\
From:     Jack Lissauer\\
To:     Darin Ragozzine \\

\noindent hi Darin,
\noindent i'll be passing the tex file back to you in a few hours.  i'd like you 
to flesh out your results and plans as much as possible, then create a 
new pdf and inform the whole ttv group.
one item you can work on now if you have time:

\smallskip
\noindent A brief discussion of why these are awesome, based on Ragozzine \& 
Holman 2010.}

\noindent A measure of our sense of urgency is that I was thinking as much of the nuclear football (the case containing  the codes required to launch nuclear warheads that is always kept near the President of the United States) as of sporting equipment.

The source code manuscript file that I sent to Darin at 11:42 was entitled `multistatistics0.17.tex'; the smallness of the version number (0.17) reflected how far I thought we were from having something suitable for refereeing 
at that time.  Even on Feb. 5, a few days after posting a draft to arXiv, the version number had only advanced to 0.48.  Version 0.54 was posted on the NX server for the \ik Science Council to review on Feb. 8.  (The NX server was our document sharing platform hosted by NASA.)  
By the evening of Feb.~23, just prior to formal submission of the manuscript to the {\it Astrophysical Journal}, we had reached version 0.80 (although some intermediate numbers may have been skipped).

During the first few years of the mission, papers written by \ik team members using data on which the project had proprietary rights needed to be submitted to the \ik Science Council (KSC) for approval.  I sent them the following message on the Sunday prior to the February 2011 \ik data release: 

\smallskip
\noindent{\bf 1/30/11, 8:16 AM: Email from JJL to KSC:}
\\
{\tt \footnotesize 
Subject:     draft paper for review\\

\noindent Dear KSC,

\noindent I hereby submit for your consideration a draft manuscript on the 
Architecture of Planetary Systems.  We would like to post the draft on 
astro-ph on Feb. 2, to establish priority for the Kepler project on new 
and important findings, and submit the paper to Ap.J. either at that 
time or, more likely, a few days to a week later.  The manuscript file 
may be downloaded from NX at:

\noindent https://nx.arc.nasa.gov/nx/dsweb/View/Collection-95014
At present, that is v0.31).
Authorship: about 90\% of the work has been done by myself, Darin 
Ragozzine, and Dan Fabrycky, and I request that we be listed as the 
first three authors, in the order given.  The other listed authors have 
made lesser direct contributions to this work and/or major contributions 
to the Borucki et al. data release paper on which this paper is based.  
Everyone who has signed the wiki that has been up all month is included 
on the author list.  I hope that other major direct contributors to the 
Borucki et al. data release paper are sufficiently interested in this 
work to ask to be added to the author list (and to read and comment on 
the draft!); many members of that group are on the KSC mailing list, and 
I'm also sending this message to some others who aren't.  The list isn't 
complete, but unlike with Nature, more authors will be easy to add 
subsequent to submission for review by the journal.  At this point, I've 
just listed authors beyond the first 3 as a single alphabetized list, 
but if the numbers grow substantially, it might be appropriate to split 
it into two alphabetized lists based upon level of contribution.
I hope that you will be able to provide me with a response no later than 
noon on Monday, as I will be absorbed by the press event at HQ on 
Tuesday and Wednesday.

\noindent Sincerely,

\noindent jack}

\noindent As planned, Darin posted the first version of Arch I to the arXiv on Feb. 2.

\subsection{Revisions, Enhancements, Improvements}

Four versions of Architectures I were ultimately posted to arXiv.org.  The chronology and some details about these postings are given below. 
For v2, two authors were indeed added to the alphabetized list for their indirect contributions related to ground-based observing support for the planet candidate catalog  \cite{Borucki:2011} and the author order was changed, bringing forward several additional significant direct contributors to the paper and placing their names after the three primary contributors but ahead of the alphabetized list of people granted co-author status primarily for their overall \ik mission and/or to the \citet{Borucki:2011} planet candidate catalog that provided the primary data for Architectures I.
Josh Carter was added to author list for v3 because, on 24 Feb., he provided important information on the period of planet candidate KOI-730.03 (Section \ref{sec:730}).  No changes were made in the author list/author ordering subsequent to that time.  The submission chronology was as follows:\\

\noindent {\tt \footnotesize From: Darin Ragozzine\\ 
$[$v1$]$ Wed, 2 Feb 2011 20:10:33 UTC (706 KB)\\
Comments:	46 pages, 13 figures. This is a preliminary draft, some numbers may change slightly in the submitted version}
\\
Above the title of the paper, the manuscript states: \\
{\tt \footnotesize Preliminary draft. To be submitted to ApJ.\\}

{\tt \footnotesize $[$v2$]$ Thu, 24 Feb 2011 06:05:28 UTC (582 KB)\\
Comments:	58 pages, 19 figures. Submitted to ApJ\\

$[$v3$]$ Fri, 25 Feb 2011 20:54:46 UTC (586 KB)\\  
Comments:	58 pages, 19 figures. This is the version submitted to ApJ (arXiv v2 was an intermediate version despite it being listed as "submitted to ApJ")}\\
Above the title of the paper, the manuscript states:\\
{\tt \footnotesize Submitted to ApJ. Note: statistical studies will be re-done once
the Borucki et al. data paper is accepted for publication. Some
numerical results are thus expected to change, but only slightly}\\
The same statement appeared above the title in version 2, as we had planned to submit that version to the journal, but were delayed by a day to incorporate important new knowledge about KOI-730 that is discussed below. \\

{\tt \footnotesize $[$v4$]$ Fri, 5 Aug 2011 20:58:44 UTC (508 KB)\\
Comments:	27 pages, 19 figures, 8 tables, emulateapj style. Accepted to ApJ, to appear in ApJS, November 2011 issue. This version includes several minor changes to the text\\}
Note that the change in format reduced the number of pages substantially from the previous versions, despite an increase in content.\\

The Astrophysical Journal refereed paper gives the following chronology: Received 2011 February 24; accepted 2011 July 20; published 2011 October 13.

Because the paper was quite long, the refereeing process took over two months.  We dubbed the version submitted to the journal v1.00 and continued to work on the paper while it was being refereed, albeit at a much slower pace. I emailed v1.05 to Darin on 20 April. In early May, I received the following good news:

\smallskip
\noindent{\bf  Excerpts from an email from the editor in charge of the manuscript to JJL, 5/3/11, 9:22 AM:}
\\
{\tt \footnotesize 
From: rasio@northwestern.edu\\
Subject:     Your ApJ submission ApJ83041\\
...\\
Re: Architecture and Dynamics of Kepler's Candidate Multiple Transiting Planet Systems, ApJ83041\\
... the referee thinks highly of your work and has only a few suggestions for relatively minor changes...\\
\smallskip
\noindent Frederic A. Rasio\\
Scientific Editor\\
The Astrophysical Journal\\
rasio@northwestern.edu\\             
===================================================\\
\noindent Referee's Report:\\

\noindent Report on "Architecture and Dynamics of Kepler's Candidate Multiple Transiting Planet Systems" by Lissauer et al.\\

\noindent This paper is a first cut discussion of the statistics of the candidate multiple planet systems discovered by Kepler. Among the many important results presented are: (1) large planets R$>$R\_neptune have a much lower incidence of multiplicity than smaller radii planets, (2) in 2 planet systems, the radii (and therefore presumably masses) tend to be approximately equal, (3) remarkably, numerical integrations of the orbits show that almost all of the candidate systems are dynamically stable, strongly suggesting that a large fraction of them are real, (4) $~$20\% of the multiple systems are close to 2:1 commensurability with a preference to be just outside rather than inside, (5) the multiplicity frequencies imply a low inclination dispersion. These results have important implications for planet formation and evolution, and this paper should certainly be published in ApJ.\\}

\noindent The reviewer then offered ten minor, but in most cases helpful, comments, the most significant of which motivated us to improve upon the method by which we tested the G statistic.

{\tt \footnotesize 





 
\noindent Equation 8 defining the G statistic:  give a reference for this statistic. Why is a statistic that gives no weight to zero expected or observed objects preferred here?\\



}
\noindent We responded to this comment by opting to use a different statistical test.  

The most substantitive changes to the manuscript were at our own initiative rather than in response to the referee report. We added two new subsections (5.6 and 5.7) discussing four systems with particularly interesting period ratios and revised our discussion of KOI-730 (Section \ref{sec:730}).  The most fundamental change was to augment our analysis of the coplanarity of planetary systems by including a model in which the inclinations of planets within a given planetary system were Rayleigh distributed, but the mean of this distribution was not the same for all systems but rather was itself Rayleigh distributed.  Unlike the model with the inclinations of planets in all systems being drawn from the same Rayleigh distribution, this ``Rayleigh of Rayleighs'' allowed us to fit the entire set of (non-giant) \ik planet candidates, including systems with only one transiting planet, using a single population.   

Just a few days after submitting the revised version of the manuscript, I received the acceptance letter (which did not request any changes) on July 20, the 44$^{th}$ anniversary of Apollo 11's landing on the Moon.  Proofs of the typeset article arrived on August 8.  The article was published in November, as one of 14 \ik research papers comprising a special issue of the {\it Astrophysical Journal Supplement} that I organized.  The \ik project purchased hundreds of copies of this issue and distributed them to scientists at various conferences.

\subsection{KOI-730}\label{sec:730}

The \citet{Borucki:2011} catalog listed four planet candidates for KOI-730. The orbital period of planet candidate KOI-730.03 was given as  9.86 days. As noted in the first two versions of Arch I posted on arXiv, this placed it in a co-orbital configuration with KOI-730.02, whose orbital period is 9.85 days. (Note that the small difference in periods between the two planets was consistent with the large-amplitude of libration around the triangular Lagrangian points required by the 118$^\circ$ separation in longitude between the two planets at the start of \ik mission according to the then-measured periods and phases of the planets.)    

On Feb. 24, Josh Carter provided us with strong evidence that planet candidate KOI-730.03's correct orbital period is 19.72 days rather than 9.86 days; this placed KOI-730.02 and KOI-730.03 in 2:1 resonance.  As neither of these two planet candidates was, on its own, unusual in terms of size or orbital period, the difference was minor for the Borucki et al.~catalog paper, where it was never changed (it was, however, changed in subsequent catalogs released by the project).  But it was significant for Arch I, which devoted an entire subsection to KOI-730.  

In the first three versions of the paper posted to arXiv and the one submitted to the journal in February, the subsection began ``While few nearly exact mean motion resonances are evident in the sample of \ik planetary
candidates, one system stands out as exceptional: the periods of the four candidates in KOI-730 satisfy the ratio 6:4:4:3 to $\sim$ 1 part in 1000 or better. This system is the first to show
evidence for extrasolar ``Trojan'' or co-orbital planets, which have been suggested to be theoretically
possible ...''.  In v3 and the manuscript submitted to the {\it Astrophysical Journal}, we added the following qualifier later in the subsection ``An alternative interpretation of the data is that the period of
KOI-730.03 is twice the nominal period, giving a period ratio of the candidates as 8:6:4:3.'' The final version of the paper, both published in the journal and on arXiv, gives the period ratio of the candidates as 8:6:4:3, but notes the initial misinterpretation of the data.

\subsection{Citations}
The first two Architectures papers have garnered substantial attention within the exoplanet community; \cite{Lissauer:2011a} is listed on ADS as having been cited 407 times as of 28 January 2019  and \cite{Fabrycky:2012a} has 344 cites. The third and perhaps final paper in the series is currently being written.

\section{Discussion}

As researchers work to generalize the theory of planet formation and dynamical evolution to account for both exoplanets and our own Solar System, we must have a clear picture of the properties of planets and the systems in which they reside.  This need goes far beyond occurrence rates, planetary bulk properties, and atmospheric chemical abundance measurements, all of which provide important, but limited, information.  Rather, it requires a comprehensive knowledge of the relationships between planets within a given system such as  correlations in planet sizes and masses, orbital period ratios, eccentricities and inclinations, along with how these quantities vary with the properties of the stellar host.

The \kepler\ data are a significant trove of this kind of information.  The more than 600 multiplanet systems that \ik discovered (despite the added difficulty in finding additional planets in a \ik light curves after the transits of the planet with the strongest signal have been removed---see \citealt{Zink:2019} for a quantitative discussion) are particularly useful since they provide intra-system information. Many of the multi-transiting systems with TTVs allow us to measure the masses of planets that would otherwise be elusive---either because the planets are too small, or the stars are too dim.  Indeed, a large fraction of the highest quality mass/radius measurements for smaller planets with orbital periods of more than a few days came through a TTV analysis in a multiplanet system.  The two papers whose histories we have focused on herein pioneered new territory in the study of exoplanets and exoplanetary systems, paving way for new lines of research in these systems.  

\section*{Acknowledgements}
JHS and JJL are supported by NASA grant NNX17AK94G.  We thank the historical editor Joann Eisberg for suggesting revisions that improved the manuscript.

\section*{References}


\end{document}